\documentstyle[12pt,a4,psfig]{article}
\newcommand\textfrac[2]{{\textstyle\frac{#1}{#2}}}
\def\delb{\mbox{$\bar{\delta}_{b}$}}
\def\vsk#1{\noalign{\vskip#1 cm}}
\def\vsp#1{\vspace{#1 cm}}
\def\ov{\overline}

\def\sm{${\rm SU(2)}_L\times {\rm U(1)}_Y$\ }
\def\gm5{\gamma_5}
\def\smw{{\mbox{\scriptsize `SM'}}}
\def\smr{{\rm SM}}
\def\mmz{m_Z^2}
\def\hph{\hphantom{-}}
\def\lenc{L.E.N.C.\ }

\def\mt{m_t^{}}
\def\mh{m_H^{}}
\def\xt{x_t^{}}
\def\xh{x_H^{}}
\def\mh{m_H^{}}

\def\msbar{$\ov{{\rm MS}}$}
\def\ds{\Delta S}
\def\dt{\Delta T}
\def\ebar{\bar{e}^2}
\def\sbar{\bar{s}^2}
\def\cbar{\bar{c}^2}
\def\gzbar{\bar{g}_Z^2}

\def\abar{\bar{\alpha}}
\def\ehat{\hat{e}}
\def\shat{\hat{s}}
\def\chat{\hat{c}}
\def\gzhat{\hat{g}_Z}
\def\ghat{\hat{g}}

\def\etal{{\it et al.}}
\newcommand{\beq}{\begin{equation}}
\newcommand{\eeq}{\end{equation}}
\newcommand{\bea}{\begin{eqnarray}}
\newcommand{\eea}{\end{eqnarray}}
\newcommand{\bsub}{\begin{subequations}}
\newcommand{\esub}{\end{subequations}}

\renewcommand{\theequation}{\thesection.\arabic{equation}}
\newcommand{\clean}{\setcounter{equation}{0}}

\def\PRD#1#2#3{Phys. Rev. {\bf D#1} (19#2) #3}
\def\NPB#1#2#3{Nucl. Phys. {\bf B#1} (19#2) #3}
\def\ZPC#1#2#3{Z. Phys. {\bf C#1} (19#2) #3}
\def\PLB#1#2#3{Phys. Lett. {\bf B#1} (19#2) #3}
\def\PRL#1#2#3{Phys. Rev. Lett. {\bf #1} (19#2) #3}
\def\MPL#1#2#3{Mod. Phys. Lett. {\bf A#1} (19#2) #3}
\makeatletter
\newtoks\@stequation

\def\subequations{\refstepcounter{equation}%
  \edef\@savedequation{\the\c@equation}%
  \@stequation=\expandafter{\theequation}
  \edef\@savedtheequation{\the\@stequation}
  \edef\oldtheequation{\theequation}%
  \setcounter{equation}{0}%
  \def\theequation{\oldtheequation\alph{equation}}}

\def\endsubequations{%
  \ifnum\c@equation < 2 \@warning{Only \the\c@equation\space subequation
    used in equation \@savedequation}\fi
  \setcounter{equation}{\@savedequation}%
  \@stequation=\expandafter{\@savedtheequation}%
  \edef\theequation{\the\@stequation}%
  \global\@ignoretrue}

\def\eqnarray{\stepcounter{equation}\let\@currentlabel\theequation
\global\@eqnswtrue\m@th
\global\@eqcnt\z@\tabskip\@centering\let\\\@eqncr
$$\halign to\displaywidth\bgroup\@eqnsel\hskip\@centering
     $\displaystyle\tabskip\z@{##}$&\global\@eqcnt\@ne
      \hfil$\;{##}\;$\hfil
     &\global\@eqcnt\tw@ $\displaystyle\tabskip\z@{##}$\hfil
   \tabskip\@centering&\llap{##}\tabskip\z@\cr}

\makeatother

\begin{document}
\thispagestyle{empty}
\vspace*{-15mm}
\baselineskip 10pt
\begin{flushright}
\begin{tabular}{l}
{\bf KEK-TH-521}\\
{\bf hep-ph/9707334}\\
{\bf July 1997}
\end{tabular}
\end{flushright}
\baselineskip 18pt 
\vglue 15mm 

\begin{center}
{\Large\bf
Constraints on four-Fermi contact interactions 
from low-energy electroweak experiments
}
\vspace{5mm}

\def\thefootnote{\alph{footnote}}
\setcounter{footnote}{0}
{\bf
Gi-Chol Cho$^{1,}$\footnote{Research Fellow of the Japan Society 
for the Promotion of Science}, 
Kaoru Hagiwara$^{1,2}$ and 
Seiji Matsumoto$^{1,a,}$\footnote{Present address: 
Theory Division, CERN, CH-1211, Gen\`{e}ve 23, Switzerland}
}
\vspace{5mm}
\def\thefootnote{\arabic{footnote}}
\setcounter{footnote}{0}

$^1${\it Theory Group, KEK, Tsukuba, Ibaraki 305, Japan}\\
$^2${\it ICEPP, University of Tokyo, Hongo, Bunkyo-ku, 
Tokyo 113, Japan}

\vspace{20mm}
\end{center}


\begin{center}
{\bf Abstract}\\[10mm]
\begin{minipage}{12cm}
\noindent

	We investigate the constraints on four-Fermi contact 
interactions from low-energy lepton-quark and lepton-lepton 
scattering experiments --- polarization asymmetries in electron 
(muon)-nucleon scattering 
experiments, cesium and thallium atom parity violation measurements, 
neutrino-nuclei and neutrino-electron scattering experiments.
	These constraints are then combined by assuming the lepton 
and quark universalities and \sm gauge invariance of the contact 
interaction, which leave independent six lepton-quark and three 
pure-leptonic interactions.
	Impacts of these constraints on models with an additional 
$Z$-boson are briefly discussed.
   We also present updates of the low-energy constraints 
on the $S$ and $T$ parameters.

\end{minipage}
\end{center}

\newpage
\section{Introduction} 
	Searching for an evidence of physics beyond the Standard Model 
(SM) is one of the most important subjects in the field of particle 
physics. 
	Up to now, no particles other than the SM particles have been 
found at collider experiments. 
	This fact implies that the mass scale of new particles should 
be larger than several hundred GeV. 

	Low-energy neutral current (L.E.N.C.) phenomena is mediated 
by exchange of the photon and the $Z$-boson in the SM. 
	We can parametrize new physics contributions to L.E.N.C. 
phenomenon as effective four-Fermi contact interactions.

	Generally, the effective contact interactions for neutral 
currents among quarks and leptons can be parametrized as 
\beq
{\cal L}_{NC} = \sum_{f,f'} \sum_{\alpha, \beta}\ 
\eta^{ff'}_{\alpha \beta}\ \ov{\psi_f} \gamma^\mu P_\alpha \psi_f\ 
\ov{\psi_{f'}} \gamma_\mu P_\beta \psi_{f'}~, 
\label{eq:contact}
\eeq
where $f,f'$ stand for lepton and quarks, $\alpha,\beta = L,R$ 
denote their chirality: $P_{L(R)} = (1-(+)\gamma_5)/2$. 
	The coefficients $\eta^{\ell q}_{\alpha \beta}$ have the 
dimension of (mass)$^{-2}$ which is often expressed 
as~\cite{eichten,PDG}, 
\beq
\eta^{ff'}_{\alpha \beta} = \frac{4\pi}{(\Lambda^{ff'+
}_{\alpha \beta})^2}\ \ {\rm or}\ \ 
- \frac{4\pi}{(\Lambda^{ff'-}_{\alpha \beta})^2}~.
\eeq
	The experimental limits of the scale $\Lambda$ are given 
for some combinations of $(f,f')$, $(\alpha,\beta)$ and the overall 
sign factor $(+,-)$~\cite{PDG}. 
	If the contact interactions are results of an exchange of 
an extra heavy neutral vector boson $Z_E$, they are given by 
\beq
\eta_{\alpha \beta}^{ff'} = -\frac{g_\alpha^f g_\beta^{f'}}{m_{Z_E}^2},
\label{eq:zprime}
\eeq
where $g_\alpha^f$ and $g_\beta^{f'}$ are the $Z_E$-boson couplings to 
$f_\alpha$ and $f'_\beta$, respectively.
	An exchange of a heavy boson in the ``$s$-channel'' of 
the processes $ff' \longrightarrow ff'$ or $f\ov{f'} \longrightarrow 
f\ov{f'}$ can also be expressed in the form of (\ref{eq:contact}) 
as long as we ignore the contribution which does not interfere with 
the leading SM amplitudes.
	In this report, we present constraints from low-energy 
electroweak measurements on the coefficients $\eta_{\alpha \beta}^{ff'}$, 
in order to take advantage of the model-independent nature of the 
parametrization (\ref{eq:contact}).
	As a simple example, the constraints on the 
$\eta_{\alpha \beta}^{ff'}$ terms are used to derive constraints 
on an extra $Z$ boson parameters.
	Recently there have been renewed interest in the possible 
existence of new interactions between quarks and leptons because of 
an excess of events at high $Q^2$ $e^+ p$ inelastic scattering at HERA, 
which was reported by H1~\cite{h1} and ZEUS~\cite{zeus} collaborations.
	Such an excess of events may be reproduced by introducing 
some lepton-quark contact interactions~\cite{hera-contact}. 
	Therefore, it is worth examining constraints on those contact 
interactions from all low-energy electroweak measurements as 
model-independent as possible.

	We study in this report the following four types of 
low-energy electroweak experiments; 
(i) polarization asymmetry of the charged lepton scattering 
off nucleus target, 
(ii) parity violation in cesium and thallium atoms, 
(iii) inelastic $\nu_\mu$ scattering off nucleus target and 
(iv) $\nu_\mu (\bar{\nu}_\mu)$ -- electron scattering experiments. 
	Individual experiment receives contributions of a different 
combination of the contact interactions. 
	We therefore present constraints on the contact terms 
from each experiment separately. 
	These constraints are then combined by assuming that the 
contact interactions satisfy the flavor universality and the 
\sm gauge invariance of the SM.

	This paper is organized as follows. 
In the next section, we review the approach of Ref.~\cite{hhkm} 
that we adopt for calculating the 
low-energy electroweak observables with the model-independent 
framework of Ref.~\cite{jekim}. 
	Section 3 is devoted to survey the experimental data 
corresponding to the above four types of experiments.  
	The SM predictions are given both as functions of $\mt$ and 
$\mh$ in the minimal SM and as a function of $S$ and $T$~\cite{stu} 
in generic \sm model.
We adopt the slightly modified version of the $S$ and $T$ variables 
that was introduced in Ref.~\cite{hhkm}. 
	In section 4 we present individual constraints on the $S$ and 
$T$ parameters in the generic \sm model, and compare the low-energy 
constraints with the constraints from the $Z$-pole experiments~\cite{hhm}. 
	They update the results of Ref.~\cite{hhkm, takeuchi,hhm}.
	In section 5 we obtain constraints on the contact interactions 
by assuming no new physics contributions to the $S$ and $T$ parameters.
	In section 6, we present constraints on models with an extra 
$Z$-boson as an example of the use of our model-independent parametrization 
of the low-energy data in terms of the four-fermi contact interactions.
\clean
\section{Framework}
The effective Lagrangian for the 
lepton-quark four-Fermi interaction is given as follows;
\begin{eqnarray}
{\cal L}_{eff}^{\ell q} &=& -\frac{G_F}{\sqrt{2}} \sum_{q=u,d} \biggl [ 
C_{1q}\ \ov{\psi_\ell} \gamma^\mu \gm5 \psi_\ell\ \ov{\psi_q} \gamma_\mu 
\psi_q \nonumber \\ 
& & + \ 
C_{2q}\ \ov{\psi_\ell} \gamma^\mu \psi_\ell\ \ov{\psi_q} \gamma_\mu\gm5 
\psi_q  + 
C_{3q}\ \ov{\psi_\ell} \gamma^\mu \gm5 \psi_\ell\ \ov{\psi_q} 
\gamma_\mu\gm5 \psi_q  
\biggr ], 
\end{eqnarray}
where $\ell = e,\mu,\tau$ and $q=u,d$.
	Adding to the model-independent parameters $C_{1q}$ and $C_{2q}$ 
of Ref.~\cite{jekim}, we introduce the parameter $C_{3q}$ as the 
coefficient of the axial vector-axial vector current. 
	They can be expressed in terms of the helicity amplitude 
$M_{\alpha \beta}^{\ell q}$ of Ref.~\cite{hhkm} as 
\bsub
\begin{eqnarray}
C_{1q} &=& \frac{1}{2\sqrt{2} G_F} \biggl( \hph 
M_{LL}^{\ell q} + 
M_{LR}^{\ell q} - M_{RL}^{\ell q} - M_{RR}^{\ell q} \biggr ), 
\\
C_{2q} &=& \frac{1}{2\sqrt{2} G_F} \biggl( \hph 
M_{LL}^{\ell q} -
M_{LR}^{\ell q} + M_{RL}^{\ell q} - M_{RR}^{\ell q} \biggr ), 
\\
C_{3q} &=& \frac{1}{2\sqrt{2} G_F} \biggl( -M_{LL}^{\ell q} + 
M_{LR}^{\ell q} + M_{RL}^{\ell q} - M_{RR}^{\ell q} \biggr ).
\end{eqnarray}
\esub
	With the above identification, $q^2$-dependence of higher order 
electroweak effects are properly taken into account by making use of 
the effective Lagrangian formalism. 
	Introducing the effective form factors $\ebar(q^2)$, $\gzbar(q^2)$, 
$\sbar(q^2)$, the helicity amplitude $M_{\alpha \beta}^{ff'}$ of the 
neutral current process $f_\alpha f'_\beta \longleftrightarrow 
f_\alpha f'_\beta$ can be expressed as~\cite{hhkm}; 
\bea
     M^{ff'}_{\alpha \beta}(q^2)^\smw &=&
     \frac{1}{q^2}
       \Bigl\{\;  (Q_f\,Q_{f'})\,\bigl[\ebar(q^2)
               +\ehat^2\,\Gamma_1^{f}(q^2) +\ehat^2\,\Gamma_1^{f'}(q^2)\bigr]
     \nonumber \\
     && \qquad
               + (Q_f\,{I_3}_{f'})\,\ehat^2\, \ov{\Gamma}_2^{f'}(q^2)
               + (Q_{f'}\,{I_3}_f)\,\ehat^2\, \ov{\Gamma}_2^f(q^2) \Bigr\}
     \nonumber \\
     &+&\frac{1}{q^2-m_Z^2}
       \Bigl\{\; ({I_3}_f -Q_f\shat^2)\,({I_3}_{f'} -Q_{f'}\shat^2)\gzbar(0)
     \nonumber \\
     && \qquad\qquad\quad -({I_3}_f -Q_f\shat^2)\, Q_{f'}\,
                        \gzhat^2\, \bigl[\sbar(q^2)-\shat^2\bigr]
     \nonumber \\
     && \qquad\qquad\quad -({I_3}_{f'} -Q_{f'}\shat^2)\, Q_f\,
                        \gzhat^2\, \bigl[\sbar(q^2)-\shat^2\bigr]\,
       \Bigr\}
     \nonumber \\
     &+&B_{ff'}^{NC}(0,0) +O\Bigl(\ehat^2\frac{q^2}{m_W^2}\Bigr)\,, 
\label{eq:amp}
\eea
in the generic \sm model, where $\ehat, \gzhat$ are the \msbar\ 
couplings that satisfy $\ehat = \ghat \shat = \gzhat \shat \chat$ 
which are renormalized as 
	$\ehat^2 = \ebar(m_Z^2)$ and $\shat^2 = 1-\chat^2 
= \sbar(m_Z^2)$.
$Q_f, I_{3f} (f=\ell, q)$ denote the electric charge and 
the third component of weak isospin, respectively, of the corresponding 
fermion $f_\alpha$. 
	$\Gamma_i^{f}, \ov{\Gamma}_i^{f}(i=1,2)$ and $B_{ff'}^{NC}$ 
stand for the vertex and box corrections, respectively, and their 
explicit forms are given in Appendix A of Ref.~\cite{hhkm}.
	In the presence of the contact interactions, the complete 
helicity amplitude is given by the sum of Eq.~(\ref{eq:amp})
and the contact term; 
\beq
M^{ff'}_{\alpha \beta}(q^2) = 
M^{ff'}_{\alpha \beta}(q^2)^\smw + \eta^{ff'}_{\alpha \beta}.
\eeq  
	Accordingly the coefficient $C_{iq}$ of the effective 
lepton-quark interactions can be divided into two pieces as, 
\beq
C_{iq} = C_{iq}^{\smw} + \Delta C_{iq}, 
\eeq
where the first term denotes the contribution with the generic 
\sm model, while the second term arises from the contact interactions. 
	Hereafter, we denote by `SM' the predictions of the generic 
\sm model, where the form factors $\gzbar(m_Z^2)$ and $\sbar(m_Z^2)$, 
or $S$ and $T$ parameters, are treated as free parameters. 
	In the minimal SM, they are determined by $\mt$ and $\mh$, 
whose explicit forms are given below~\cite{hhkm, hhm}. 
	The coefficients $C_{iq}^{\smw}$ are approximately expressed as 
\bsub
\begin{eqnarray}
C^{\smw}_{1q} &=& \hph I_{3q} - 2Q_q \sin^2\theta_W~~+~~
{\rm higher~order~terms}, \\
C^{\smw}_{2q} &=& \hph I_{3q}(1 - 4\sin^2\theta_W )~~+~~
{\rm higher~order~terms} , \\
C^{\smw}_{3q} &=& -I_{3q}
~~+~~{\rm higher~order~terms}. 
\end{eqnarray}
\esub
The terms $\Delta C_{iq}$ receive contributions from 
the contact interactions, 
\bsub
\begin{eqnarray}
\Delta C_{1q} &=& 
\frac{1}{2\sqrt{2} G_F} \biggl( \hph \eta_{LL}^{\ell q} + 
\eta_{LR}^{\ell q} - \eta_{RL}^{\ell q} - \eta_{RR}^{\ell q}
\biggr),\\
\Delta C_{2q} &=& 
\frac{1}{2\sqrt{2} G_F} \biggl( \hph \eta_{LL}^{\ell q} - 
\eta_{LR}^{\ell q} + \eta_{RL}^{\ell q} - \eta_{RR}^{\ell q}
\biggr), \\
\Delta C_{3q} &=& 
\frac{1}{2\sqrt{2} G_F} \biggl( -\eta_{LL}^{\ell q} + 
\eta_{LR}^{\ell q} + \eta_{RL}^{\ell q} - \eta_{RR}^{\ell q}
\biggr). 
\end{eqnarray}
\label{eq:c_def}
\esub
For neutrino-quark scattering, 
it is conventional to introduce the model 
independent parameters, $g_\alpha^2$ and $\delta_\alpha^2$ 
$(\alpha = L, R)$~\cite{fh}; 
\bsub
\bea
g^2_\alpha &\equiv& u_\alpha^2 + d_\alpha^2, \\
\delta^2_\alpha &\equiv& u_\alpha^2 - d_\alpha^2,
\eea
\esub
where, $u_\alpha$ and $d_\alpha$ are given by using the helicity 
amplitude as~\cite{hhkm}, 
\bsub
\bea
q_\alpha &=& -\frac{1}{2\sqrt{2}G_F} M_{L\alpha}^{\nu_\mu q} 
\\
&=& \hph q_\alpha^{\smw} + \Delta q_{\alpha}~~~~(q=u,d),
\eea
\label{eq:q_def}
\esub
\\
The `SM' contributions are obtained from Refs.~\cite{hhkm, hhm} 
and the contribution from the contact interactions is 
\beq 
\Delta q_\alpha = -\frac{1}{2\sqrt{2}G_F} \eta_{L\alpha}^{\nu_\mu q}.
\eeq
For the neutrino-electron scattering experiments, we use 
the experimental data of the total cross sections for 
$\nu_\mu$-$e$ and $\bar{\nu}_\mu$-$e$ scatterings, which are 
expressed in terms of the helicity amplitude as~\cite{hhkm};
\bsub 
\bea
\frac{\sigma^{\nu e}}{E_\nu} &=& 
\frac{m_e}{4\pi} \biggl \{ 
\biggl |
M_{LL}^{\nu_\mu e}(\langle Q^2 \rangle = m_e E_\nu  )
\biggr |^2
+ 
\frac{1}{3}
\biggl |
M_{LR}^{\nu_\mu e}(\langle Q^2 \rangle = \frac{m_e E_\nu}{2}  )
\biggr |^2
\biggr \} , 
\\
\vsk{0.3}
\frac{\sigma^{\ov{\nu} e}}{E_{\ov{\nu}}} &=& 
\frac{m_e}{4\pi} \biggl \{ 
\frac{1}{3}
\biggl |
M_{LL}^{\nu_\mu e}(\langle Q^2 \rangle = \frac{m_e E_{\ov{\nu}}}{2}  )
\biggr |^2
+ 
\biggl |
M_{LR}^{\nu_\mu e}(\langle Q^2 \rangle = m_e E_{\ov{\nu}}  )
\biggr |^2
\biggr \} .
\eea
\label{eq:ne}   
\esub
\\
Here we replace the integral over the momentum transfer 
$Q^2$ by $m_e E_\nu$. 
The amplitudes are then parametrized as 
\beq
M_{L\alpha}^{\nu_\mu e} =
(M_{L\alpha}^{\nu_\mu e})^\smw + \eta_{L\alpha}^{\nu_\mu e}.
\eeq

	Next, we briefly review how we estimate the contribution of 
generic \sm model.
	In the formalism of Ref.~\cite{hhkm}, the 
\lenc observables are expressed in terms of two form factors, 
$\gzbar(0)$ and $\sbar(0)$. 
	They can be expressed by the $S$ and $T$-parameters~\cite{stu} 
as follows;
\bsub    
\bea
\gzbar(0) &\approx& 0.5456 + 0.0040 T, 
	\label{eq:gzb0}\\
\sbar(0) &\approx& 0.2418 + 0.0034 S'' - 0.0023 T,  
\eea
\esub
where $S''$ is defined by 
\beq
S'' \equiv S - 1.30 \delta_\alpha. 
\label{eq:s_double_prime}
\eeq
	The parameter $\delta_\alpha \equiv 1/\abar(m_Z^2) - 128.72 $ 
was introduced in Ref.~\cite{hhkm} to take care of the uncertainty in 
$\abar(m_Z^2)$. 
	The recent estimation by Eidelman and Jegerlehner~\cite{eidelman} 
gives $\delta_\alpha = 0.03 \pm 0.09$~\cite{hhm}. 
	In the minimal SM, the $S$ and $T$-parameters are parametrized by 
$\mt$ and $\mh$ as~\cite{hhm}, 
\bsub
\bea
S_{\smr} &\approx& -0.233 - 0.007 \xt + 0.091 \xh - 0.010 x_H^2, \\
T_{\smr} &\approx& \hph 0.879 + (0.130-0.003\xh)\xt + 0.003 x_t^2 - 
0.079 \xh \nonumber 
\\ 
&&~~~~- 0.028 x_H^2 + 0.0026 x_H^3,
\eea
\esub
where $\xt$ and $\xh$ are 
\bsub
\bea 
\xt &\equiv& (\mt - 175\ {\rm GeV})/10\ {\rm GeV}, \\
\xh &\equiv& \log(\mh/100\ {\rm GeV}), 
\eea
\label{xt_xh}
\esub
respectively\footnote{
These parametrizations are valid in the mass range 
160 GeV $< \mt < $ 185 GeV and 40 GeV $< \mh < $ 1000 GeV.}.
	It is convenient to introduce the following parameters; 
\bsub 
\bea
\Delta S &\equiv& S - S_{SM}^0 = S + 0.233, \\
\Delta T &\equiv& T - T_{SM}^0 = T - 0.879,
\eea
\esub
where $S_{\smr}^0$ and $T_{\smr}^0$ are the SM predictions for $S$ 
and $T$ at $\mt = 175$ GeV, $\mh=100$ GeV.  

%
%
%
%
\clean
\section{Low-energy electroweak observables}
\subsection{Polarization asymmetry of charged lepton-
nucleus scattering}
	In this subsection, we study four experiments on charged 
lepton-nucleus scattering, in which two types of observables were measured. 
First, polarization asymmetry $A$ of charged lepton scattering 
off nucleus target 
\begin{equation}
A \equiv \frac{d\sigma_R - d\sigma_L}{d\sigma_R + d\sigma_L}, 
\end{equation}
has been measured in $e$D-scattering at SLAC~\cite{slac}, in $e$C 
scattering at Bates~\cite{bates} and in $e$Be scattering at 
Mainz~\cite{mainz}.
	Here $d\sigma_{L(R)}$ denotes the differential cross section 
of the left- (right-) handed lepton scattering off target.
	These are Parity-odd asymmetries.
	Another type of the asymmetry is the polarization and charge 
asymmetry parameter $B$
\begin{equation}
B \equiv \frac{d\sigma_L^+ - d\sigma_R^-}{d\sigma_L^+ + 
d\sigma_R^-}, 
\end{equation} 
where $d\sigma_{R(L)}^{-}$ is the differential cross section of right- 
(left-) handed negatively charged lepton scattering off nucleus target, 
wheres $d\sigma_{R(L)}^+$ denotes those of positively charged 
anti-lepton scattering. 
	The parameter $B$ has been measured in $\mu^{\pm}$C scattering 
at CERN~\cite{cern}. 
	The measurement of the asymmetry parameter $A$ constrains a 
combination of $C_{1q}$ and $C_{2q}$, wheres the asymmetry $B$ constrains 
the parameter $C_{3q}$.

\subsubsection{SLAC $e$D experiment}
	The historic SLAC experiment~\cite{slac} that established 
parity violation in the electron-quark neutral current scattering 
still gives non-trivial constraints on new physics contribution.
	Here we quote the results of the analysis of Ref.~\cite{hhkm}. 
	The parameter $A_{SLAC}$ is expressed in Ref.~\cite{hhkm} as 
\bea 
A_{SLAC} &=& -\frac{6G_F Q^2}{5\sqrt{2}\ebar(-Q^2)}\biggl\{ 
\biggl( 2C_{1u}-C_{1d}\biggr)
\biggl(1-\frac{3}{4}c\biggr) \nonumber \\
&&~~~+\biggl( 2C_{2u}-C_{2d}\biggr)
\biggl( b+\frac{5}{12}c\biggr) \biggr\},
\eea
where the terms $b$ and $c$ represent deviations from the valence-quark 
approximation. 
	After studying the uncertainties in the correction terms 
$b$ and $c$, the following constraints were obtained~\cite{hhkm} for the 
model-independent parameters
\bea
\begin{array}{l}
	\left.
	\begin{array}{l}
	2C_{1u}-C_{1d} =
	\hphantom{-}
	0.94 \pm 0.26 \\
	\vsk{0.2}
	2C_{2u}-C_{2d} = -0.66 \pm 1.23 
	\end{array}
	\right \}~~~~ 
	\rho_{\rm corr} = -0.975. 
\end{array}
\label{eq:ed_exp}
\eea

	The theoretical prediction within the generic \sm model at 
$\langle Q^2 \rangle \simeq 1.5\ {\rm GeV}^2$ is given as;
\bsub
\bea 
(2C_{1u}-C_{1d})^{\smw} &\approx& 0.723 - 0.0115 \Delta S + 0.0129 \Delta T, \\
(2C_{2u}-C_{2d})^{\smw} &\approx& 0.105 - 0.0207 \Delta S + 0.0144 \Delta T.
\eea
\label{eq:slac_sm1}
\esub
\\
The minimal SM predicts
\bsub
\bea 
(2C_{1u}-C_{1d})^{\smr} &\approx& 0.723 + 0.002 \xt - 0.0024 \xh, \\
(2C_{2u}-C_{2d})^{\smr} &\approx& 0.105 + 0.002 \xt - 0.0032 \xh.
\eea
\label{eq:slac_sm2}
\esub

	\subsubsection{CERN $\mu^\pm $C experiment}
	Charge and polarization asymmetry of $\mu^\pm$ deep inelastic 
scattering off $^{12}{\rm C}$ target has been measured at CERN~\cite{cern}. 
In the parton model, the asymmetry parameter $B$ is given as 
\begin{eqnarray}
B &=& -\frac{6 G_F}{5\sqrt{2} \ebar (-Q^2) }
\frac{1-(1-y)^2}
{1+(1-y)^2} Q^2 \biggl[ (-2C_{3u}^\mu + C_{3d}^\mu ) 
+ |\lambda|(-2C_{2u}^\mu + C_{2d}^\mu) \biggr] \nonumber \\
&&~~~~~~\times \frac{q(x)-\ov{q}(x)}{D(x)}, 
\label{eq:cern}
\end{eqnarray}
where $\lambda$ denotes the effective $\mu^\pm$ beam polarization. 
	Here we assumed that charm and strange quarks have the same 
effective interaction as up and down quarks, respectively: 
$C_{1c}^\mu = C_{1u}^\mu$ and $C_{1s}^\mu = C_{1d}^\mu$.
The superscript $\mu$ is put to remind us that these parameters 
are for the $\mu$-$q$ scattering. 
All the other experiments are done for electron scatterings, and 
the corresponding superscript $e$ is suppressed. 
	$q(x),\ov{q}(x)$ and $D(x)$ are expressed in terms of the quark 
and anti-quark  distribution functions in the nucleus as 
\bsub
\begin{eqnarray}
q(x) &=& u(x) + d(x) + s(x) + c(x), \\
\ov{q}(x) &=& \ov{u}(x) + \ov{d}(x) + \ov{s}(x) + \ov{c}(x), \\
D &=& u(x) + \ov{u}(x) + d(x) + \ov{d}(x) + \frac{2}{5}( s(x) 
+ \ov{s}(x) ) \nonumber \\
&&~~~~+ \frac{8}{5}(c(x) + \ov{c}(x)). 
\end{eqnarray}
\esub
The experiment has been performed at two different beam energies, 
$E=200$ GeV and 120 GeV. 
The mean momentum transfer of the experiments may be estimated 
as $\langle Q^2 \rangle \simeq 50\ {\rm GeV}^2$~\cite{souder}.   
The experimental data give the following constraints on 
the coefficients of $C_{2q}$ and $C_{3q}$; \\
(i) $E = 200$ GeV: 
\bea
\begin{array}{l}
(2C_{3u}-C_{3d}) + |\lambda|(2C_{2u} - C_{2d}) = 
-1.51 \pm 0.43, \\
\vsk{0.2}
|\lambda| = 0.81 \pm 0.04, 
\end{array}
\eea
(ii) $E = 120$ GeV: 
\bea
\begin{array}{l}
(2C_{3u}-C_{3d}) + |\lambda|(2C_{2u} - C_{2d}) =
-1.79 \pm 0.83, \\
\vsk{0.2}
|\lambda| = 0.66 \pm 0.05.
\end{array}
\eea
Combining the above two constraints, we find 
\bea
\begin{array}{l}
	\left.
	\begin{array}{l}
	2C_{3u}-C_{3d} = -3.01 \pm 4.92 \\
	\vsk{0.2}
	2C_{2u}-C_{2d} = \hph 1.85 \pm 6.31 \\
	\end{array}
	\right \}~~~~ 
	\rho_{\rm corr} = -0.997.
\end{array}
\eea
The theoretical prediction of the generic \sm model at 
$\langle Q^2 \rangle \simeq 50\ {\rm GeV}^2$ is given as;
\bsub
\bea 
(2C_{3u}-C_{3d})^{\smw} &\approx& -1.505 - 0.011 \Delta T, \\
(2C_{2u}-C_{2d})^{\smw} &\approx& \hph 0.109 - 0.021 \Delta S + 0.015 \Delta T.
\eea
\label{eq:cern_sm1}
\esub
\\
The minimal SM predicts, 
\bsub
\bea 
(2C_{3u}-C_{3d})^{\smr} &\approx& -1.505 - 0.001 \xt + 0.001 \xh, \\
(2C_{2u}-C_{2d})^{\smr} &\approx& \hph 0.109 + 0.002 \xt - 0.003 \xh.
\eea
\label{eq:cern_sm2}
\esub
\\
The superscript $\mu$ has been suppressed since the predictions 
of the SM and the generic \sm model are essentially the same 
for $e$ and $\mu$. 
The difference between the predictions for 
$2C_{2u}-C_{2d}$ in Eqs.~(\ref{eq:slac_sm1}) and (\ref{eq:cern_sm1}) 
reflects different $\langle Q^2 \rangle$ of the two experiments.
	\subsubsection{Bates $e$C experiment}
The polarization asymmetry parameter $A$ of electron 
elastic scattering off $^{12}{\rm C}$ target has been measured 
at Bates~\cite{bates} 
\begin{equation}
A_{Bates} = -3\sqrt{2}\frac{G_F Q^2}{\ebar(-Q^2) } 
\biggl( C_{1u} + C_{1d} \biggr). 
\end{equation}
Here we neglected contribution from quarks other than the 
$u$ and $d$ quarks, which have been estimated to be small 
under the condition of the experiment~\cite{beck} at a typical 
momentum transfer of $\langle Q^2 \rangle = 0.0225$ GeV$^2$.
From the experimental data 
\beq
A_{Bates} = (1.62 \pm 0.38) \times 10^{-6}, 
\eeq
we find by using $4\pi/\ebar(-0.0225\ {\rm GeV}^2) = 135.87$, 
\beq
C_{1u}+C_{1d} = -0.137 \pm 0.033.
\eeq
The corresponding theoretical prediction at 
$\langle Q^2 \rangle = 0.0225\ {\rm GeV}^2$ is found to be 
\beq
(C_{1u}+C_{1d})^{\smw} \approx 
-0.1522 - 0.0023 \ds + 0.0004 \dt,
\eeq
in the generic \sm model, and 
\beq
(C_{1u}+C_{1d})^{{\smr}} \approx -0.1522 + 0.0001\xt - 0.0002 \xh,
\eeq
in the minimal SM.

	\subsubsection{Mainz $e$Be experiment} 
At Mainz, polarization asymmetry of electron quasi-elastic scattering 
off $^9{\rm Be}$ target has been measured~\cite{mainz}.
The experiment was performed at the mean momentum transfer 
$\langle Q^2 \rangle \simeq 0.2025\ {\rm GeV}^2$. 
In this experiment, the asymmetry parameter $A_{Mainz}$ 
can be expressed by the following combination of the model 
independent parameters~\cite{mainz}; 
\beq
A_{Mainz} = -2.73 C_{1u} + 0.65 C_{1d} - 2.19 C_{2u} + 2.03 C_{2d},  
\eeq
and the experimental result is given by 
\beq
A_{Mainz} = -0.94 \pm 0.19. 
\eeq
Here the error is obtained by taking the quadratic sum of 
the theoretical and experimental ones as quoted in Ref.~\cite{mainz}.

The theoretical prediction in the generic \sm model is 
\beq
A_{Mainz}^{\smw} \approx 
-0.875 + 0.043 \ds - 0.035 \dt,
\eeq
and that of the minimal SM is, 
\beq
A_{Mainz}^{\smr} \approx -0.875 - 0.005 \xt + 0.007 \xh.
\eeq

	\subsection{Atomic Parity Violation}
The experimental results of parity violation in atoms are often given 
in terms of the weak charge $Q_W(A,Z)$ of the nuclei. 
Using the model-independent parameter $C_{1q}$, the weak charge of a 
nuclei ($A,Z)$ can be expressed as; 
\beq
Q_W(A,Z) 
= 2Z C_{1p} + 2(A-Z) C_{1n}, 
\eeq
where two coefficients $C_{1p}$ and $C_{1n}$ are given in terms of 
$C_{1u}$ and $C_{1d}$ as~\cite{hhkm}
\bsub
\bea
C_{1p} &\approx& 2 C_{1u} + C_{1d} + 0.01349, \\
C_{1n} &\approx& C_{1u} + 2 C_{1d} + 0.00669.
\eea
\esub
The theoretical prediction for $C_{1u}$ and $C_{1d}$ is found to be 
\bsub 
\bea
C_{1u}^{\smw} &\approx& \hph 0.18185 - 0.0045 \ds + 0.0043 \dt,  \\
C_{1d}^{\smw} &\approx& -0.34116 + 0.0023 \ds - 0.0040 \dt, 
\eea
\esub
in the generic \sm model, and 
\bsub
\bea
C_{1u}^{\smr} &\approx& \hph 0.18185 + 0.00059 \xt - 0.00073 \xh, \\
C_{1d}^{\smr} &\approx& -0.34116 - 0.00054 \xt + 0.00051 \xh, 
\eea
\esub
in the minimal SM. 

There are two accurate measurements of the weak charge 
in the cesium atom~\cite{noecker, wood};
\begin{eqnarray}
Q_W(^{133}_{55}Cs) = \left \{  \begin{array}{ll}
-71.04 \pm 1.58 \pm 0.88&~~~~~~~~~\cite{noecker},\\  
-72.11 \pm 0.27 \pm 0.88&~~~~~~~~~\cite{wood}, 
\end{array}
\right.
\end{eqnarray}
where the first error is experimental while the second 
one is theoretical. 
Both theoretical errors come from the same estimation~\cite{bllundell}, 
and hence they are 100\% correlated. 
By combining these two data, we obtain 
\beq
Q_W (^{133}_{55}Cs) 
= -72.08 \pm 0.92.
\eeq

APV experiments in the thallium atom have been performed by 
Seattle~\cite{seattle} and Oxford~\cite{oxford} groups.
The results are; 
\begin{eqnarray}
Q_W(^{205}_{81}Tl) = \left \{  \begin{array}{ll}
-114.2 \pm 1.3 \pm 4.0&~~~~~~~~~\cite{seattle},\\
-120.5 \pm 3.5 \pm 4.0&~~~~~~~~~\cite{oxford}.
\end{array}
\right.
\end{eqnarray}
The result in the second line is obtained in Ref.~\cite{rosner} 
where $Q_W$ was extracted from the report~\cite{oxford} of the 
Oxford group. 
In Ref.~\cite{seattle}, a part of the theoretical error that has 
been accounted for in Ref.~\cite{oxford} was not included.
Therefore we replaced the theoretical error of Ref.~\cite{seattle}
by $\pm 4.0$, following Ref.~\cite{rosner}. 
The two common theoretical errors are again 100\% correlated.
The combined result is given by 
\beq
Q_W(^{205}_{81}Tl) = -115.0 \pm 4.2.
\eeq

Theoretical predictions for the weak charges of $Cs$ and $Tl$ are 
\bsub
\bea
Q_W^{\smw} (^{133}_{55}Cs) &\approx& 
-73.07 - 0.745 \ds - 0.054 \dt, \\
Q_W^{\smw} (^{205}_{81}Tl) &\approx& 
-116.6 - 1.10 \ds -0.15 \dt, 
\eea
\esub
in the generic \sm model. 
In the minimal SM, they are given as; 
\bsub
\bea
Q_W^{\smr}(^{133}_{55}Cs) &\approx& 
-73.07 - 0.003 \xt - 0.062 \xh + 0.007 x_H^2, \\
Q_W^{\smr}(^{205}_{81}Tl) &\approx& 
-116.6 - 0.01 \xt - 0.09 \xh + 0.01 x_H^2.
\eea
\esub

	\subsection{$\nu_\mu$-quark scattering}
For neutrino-quark scattering, the experimental results are 
presented by using the model-independent parameters $g_\alpha^2$ 
and $\delta_\alpha^2$ (or equivalently $q_\alpha$) of Ref.~\cite{fh}. 
We examine two independent sets of experimental data; the results 
of all the old neutrino-nucleus scattering experiments as 
summarized by Fogli and Haidt (FH)~\cite{fh} and the recent 
CCFR measurement~\cite{ccfr}. 

The result of Ref.~\cite{fh} can be expressed as~\cite{hhkm}
\bea
\begin{array}{l} \left.
	\begin{array}{l}
	g_L^2 = \hph 0.2980 \pm 0.0044 \\
	g_R^2 = \hph 0.0307 \pm 0.0047 \\
	\delta_L^2 = -0.0589 \pm 0.0237 \\
	\delta_R^2 = \hph 0.0206 \pm 0.0160
	\end{array}
	\right \}~~~~ 
	\rho_{\rm corr} = \left ( 
	\begin{array}{rrrr}
	1 & -0.559 & -0.163 & 0.162 \\
	  &      1 & 0.156  & -0.037 \\
	  &        &     1  & -0.447 \\
	  &        &        &      1 
		\end{array} \right ).
\end{array}
\label{eq:fh_data}
\eea
Typical momentum transfer of these measurements is 
$\langle Q^2 \rangle_{\rm FH} = 20\ {\rm GeV^2}$.

The experiment of CCFR collaboration has been done at slightly 
higher momentum transfer 
$\langle Q^2 \rangle_{\rm CCFR} =35\ {\rm GeV^2}$. 
The result was given for the following combination of 
the model-independent parameters~\cite{ccfr}; 
\bsub
\bea
K 
&=& 1.7897 g_L^2 + 1.1479 g_R^2 - 0.0916 \delta_L^2 
- 0.0782 \delta_R^2,\\ 
&=& 0.5820 \pm 0.0049.
\eea
\label{eq:ccfr_data}
\esub

Taking into account of the difference of the typical momentum 
transfer of the two data sets, we find the following theoretical 
predictions for the model-independent parameters;
\bsub
\bea
u_L^{\smw} &\approx& 
	\biggl (
	\begin{array}{l}
	\hph 0.3465 \\
	\hph 0.3468
	\end{array}
	\biggr )
	-0.0023 \ds + 0.0041 \dt, \\
u_R^{\smw} &\approx& 
	\biggl (
	\begin{array}{l}
	-0.1549 \\
	-0.1549
	\end{array}
	\biggr )
	-0.0023 \ds + 0.0004 \dt, \\
d_L^{\smw} &\approx&
	\biggl (
	\begin{array}{l}
	-0.4296 \\
	-0.4299
	\end{array}
	\biggr )
	+ 0.0012 \ds -0.0039 \dt, \\
d_R^{\smw} &\approx& 
	\biggl (
	\begin{array}{l}
	\hph 0.0776 \\
	\hph 0.0775
	\end{array}
	\biggr )
	+ 0.0012 \ds -0.0002 \dt,
\eea
\esub
in the generic \sm models, and 
\bsub
\bea
u_L^{\smr} &\approx& 
	\biggl (
	\begin{array}{l}
	\hph 0.3465 \\
	\hph 0.3468
	\end{array}
	\biggr )
	+ 0.0005 \xt - 0.00065 \xh, \\
u_R^{\smr} &\approx& 
	\biggl (
	\begin{array}{l}
	-0.1549 \\
	-0.1549
	\end{array}
	\biggr )
	+ 0.0001 \xt - 0.00022 \xh,\\
d_L^{\smr} &\approx&
	\biggl (
	\begin{array}{l}
	-0.4296 \\
	-0.4299
	\end{array}
	\biggr )
	- 0.0005 \xt + 0.00055 \xh,\\
d_R^{\smr} &\approx& 
	\biggl (
	\begin{array}{l}
	\hph 0.0776 \\
	\hph 0.0775
	\end{array}
	\biggr )  
	\hphantom{-0.00005 \xt} + 0.00011 \xh, 
\eea
\esub
in the minimal SM. 
In the above, 
the upper and the lower numbers in each column are 
the predictions at $\langle Q^2 \rangle_{\rm FH} = 20\  
{\rm GeV^2}$ and at $\langle Q^2 \rangle_{\rm CCFR} = 35\  
{\rm GeV^2}$, respectively.
The $Q^2$-dependences of the theoretical predictions 
turned out to be negligibly small.

	\subsection{$\nu_\mu$-electron scattering}
There are three experimental results for the $\nu_\mu$-electron 
scattering~\cite{numu}. 
Here we use the combined data which are expressed in terms of 
the cross sections~\cite{hhkm};  
\bea
\begin{array}{l}
	\left.
	\begin{array}{l}
	\bigl[ \sigma^{\nu e}/E_\nu \bigr]
	(10^{-42}{\rm cm^2 / GeV})
	= 1.56 \pm 0.10, \\
	\vsk{0.2}
	\bigl[ \sigma^{\ov{\nu} e}/E_{\ov{\nu}} \bigr]
	(10^{-42}{\rm cm^2 / GeV})
	= 1.36 \pm 0.09,
	\end{array}
	\right \}~~~
	\rho_{\rm corr} = 0.51.
\end{array}
\label{eq:nu_e}
\eea
As seen from Eq.~(\ref{eq:ne}), the above 
cross sections expressed in terms of the helicity amplitudes, 
$M_{L \alpha}^{\nu_\mu e}$. 
If we ignore the $Q^2$-dependence of their matrix elements, the 
experimental data (\ref{eq:nu_e}) can be expressed in terms of 
the two independent helicity amplitudes as 
\bea
\begin{array}{l}
	\left.
	\begin{array}{l}
	M_{LL}^{\nu_\mu e} 
	= \hph 8.87 \pm 0.35, \\
	\vsk{0.2}
	M_{LR}^{\nu_\mu e} 
	= -7.73 \pm 0.36, 
	\end{array}
	\right \}~~~
	\rho_{\rm corr} = 0.13, 
\end{array}
\eea
in units of TeV$^{-2}$. 
The theoretical predictions in the generic \sm models are 
\bsub
\bea
(M_{LL}^{\nu_\mu e})^{\smw} &\approx& 
	\hph 9.02 
	-0.11 \ds + 0.14 \dt , \\
(M_{LR}^{\nu_\mu e})^{\smw} &\approx& 
	-7.70
	-0.11 \ds + 0.02 \dt, 
\eea
\esub
and those within the minimal SM are, 
\bsub
\bea
(M_{LL}^{\nu_\mu e})^{\rm SM} &\approx& 
	\hph 9.02 
	+ 0.02\xt - 0.025 \xh, \\
(M_{LR}^{\nu_\mu e})^{\rm SM} &\approx& 
	-7.70
	+ 0.003 \xt - 0.011 \xh,
\eea
\esub
all in units of TeV$^{-2}$.
The $\langle Q^2 \rangle$-dependence of the theoretical predictions 
of the matrix elements $M_{L\alpha}^{\nu_\mu e}(\langle Q^2 \rangle)$ 
has indeed been found to be negligible between 
$\langle Q^2 \rangle = m_e E_\nu$ and 
$\langle Q^2 \rangle = m_e E_\nu/2$ in Eq.~(\ref{eq:ne}), 
for $E_\nu = 25.7$ GeV (CHARM II~\cite{numu}).

\clean
\section{ Results on generic SU(2)$_L \times$U(1)$_Y$ model}
\subsection{ Constraints on $S,T$ parameters  }

Here, we study the constraints on $S$ and $T$ parameters in the 
generic \sm model from the low-energy electroweak experiments listed 
in the previous section. 
(i){\it  
Asymmetries in $\ell$-$q$ scatterings experiments;}\\
We combine the results of the four experiments, SLAC ($e$D), CERN 
($\mu^{\pm}$C), Bates ($e$C), Mainz ($e$Be) and find
\beq
S''=  -3.75  + 0.93 T   \pm 2.80.
\label{eq:eq_scatt}
\eeq
(ii){\it APV experiments;}\\
The combined result of parity violation experiments in the 
cesium and thallium atoms is given by 
\beq
S''= -1.6 - 0.063 T   \pm 1.2. 
\label{eq:apv_sum}
\eeq
As emphasized in Ref.~\cite{rosner, mr90}, 
the APV experiments constrain mainly the S parameter. 
(iii) {\it   $\nu_\mu$-$q$ scatterings;}\\
From the two data sets of the $\nu_\mu$-$q$ scattering experiments 
given in section 3.3, the following constraint is obtained, 
\bea
\begin{array}{l}
	\left.
	\begin{array}{l}
	S'' = 0.95 \pm 5.01\\
	\vsk{0.2}
	T\hphantom{''}  = 0.75 \pm 1.78
	\end{array}
	\right \}~~~~ 
	\rho_{\rm corr} = 0.979. 
\end{array}
\label{eq:stu_nu-q}
\eea
Because of the strong positive correlation between the errors, 
only the following combination is effectively constrained; 
\beq
T = 0.42 + 0.35 S'' \pm 0.36.
\eeq
(iv) {\it   $\nu_\mu$-$e$ scatterings;}\\
From the $\nu_\mu$-$e$ scattering data in section 3.4, we find
\bea
\begin{array}{l}
	\left.
	\begin{array}{l}
	S'' = 0.035 \pm 3.6 \\
	\vsk{0.2}
	T\hphantom{''}  = 0.064 \pm 3.9
	\end{array}
	\right \}~~~~ 
	\rho_{\rm corr} = 0.76. \\
\vsk{0.3} 
\label{eq:nq_ne}
\end{array}
\eea

\begin{figure}[t]
\begin{center}
\leavevmode\psfig{figure=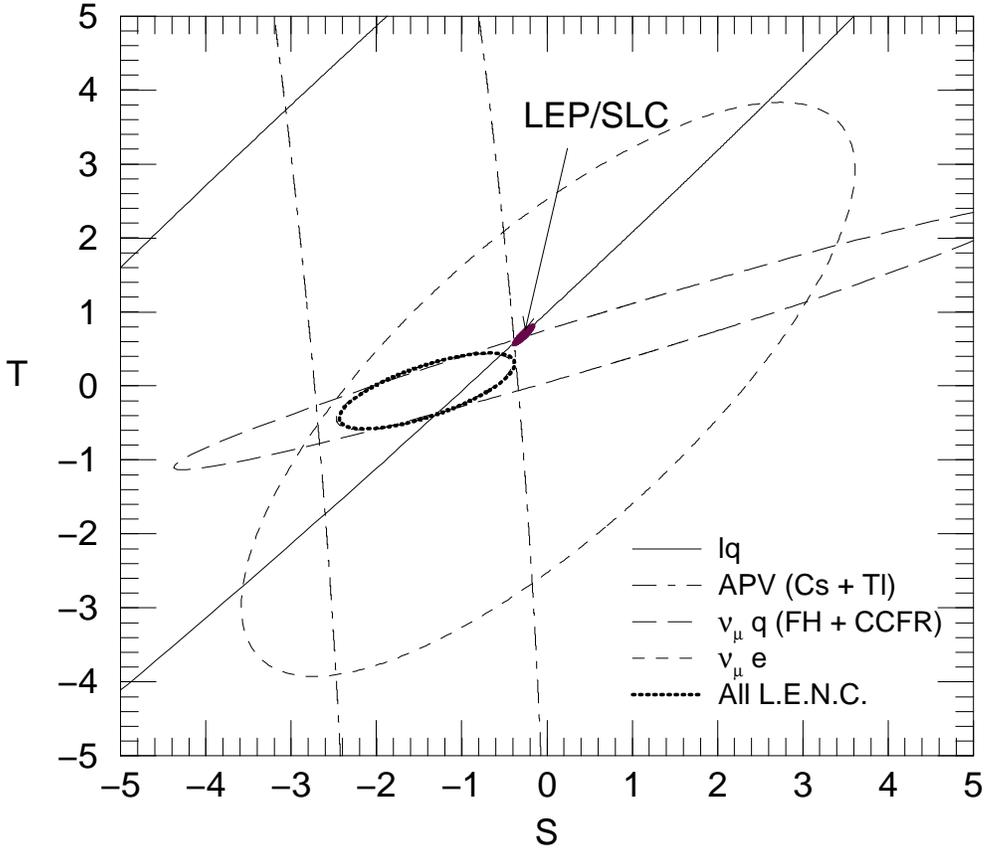,width=13cm}
\end{center}
\caption{
Fit to \lenc data on the ($S,T$) plane. 
The 1-$\sigma$ (39\% CL) allowed ranges are shown separately 
for charged lepton-quark scattering experiments, atomic 
parity violation, $\nu_\mu$-quark scattering and $\nu_\mu$-$e$ 
scattering experiments. 
Also shown are the 1-$\sigma$ allowed region of all the \lenc 
experiments as well as that of the LEP/SLC $Z$-pole measurements.}
\label{stu_fit}
\end{figure}

Summing up all the \lenc experiments, 
we find the following constraints on the $S$ and $T$ parameters, 
\bea
\begin{array}{l}
	\left.
	\begin{array}{l}
	S'' = -1.44 \pm 1.03\\
	\vsk{0.2}
	T\hphantom{''} = -0.06 \pm 0.51
	\end{array}
	\right \}~~~~ 
	\rho_{\rm corr} = 0.715, \\
\vsk{0.3}
~~~\chi^2_{\rm min}/({\rm d.o.f}) = 3.38/(13).
\end{array}
\label{eq:sum_lenc}
\eea
We show in Fig.~\ref{stu_fit} 
our results of the individual constraints, 
(\ref{eq:eq_scatt}), (\ref{eq:apv_sum}), (\ref{eq:stu_nu-q}) and 
(\ref{eq:nq_ne}) as well as the combined result (\ref{eq:sum_lenc}) 
for $\delta_\alpha = 0.03$~\cite{eidelman}. 
When compared with corresponding result of Refs.~\cite{takeuchi,hhm}, 
significant improvements are found in the constraints from the 
APV and the $\nu_\mu$-$q$ scattering experiments. 
Constraints from the $\ell$-$q$ scattering experiments are only slightly 
improved by including the CERN $\mu^{\pm}$C, Bates $e$C and Mainz 
$e$Be experiments in the analysis.

\subsection{Comparison with LEP/SLC data}
From the updated $Z$ shape parameter measurements by LEP/SLC, the 
effective charges $\sbar(m_Z^2)$ and $\gzbar(m_Z^2)$ have been 
extracted in Ref.~\cite{hhm}. 
It is assumed that the vertex functions beside the 
$Z b_L b_L$ vertex function $\delb (m_Z^2)$ are dominated by the SM 
contributions and the following combination~\cite{hhkm,hhm}
\beq
\alpha_s' = \alpha_s(m_Z^2)_{\ov{\rm MS}}
+ 1.54 [ \delb(m_Z^2) + 0.00995 ]
\eeq
is constrained by the hadronic decay width of the $Z$-boson.
Then, by taking $\alpha_s'$ and $\delb(m_Z^2)$ as external parameters, 
the following result has been found in Ref.~\cite{hhm};
  \begin{subequations}
   \label{fitofgzbsb}
  \begin{eqnarray}
     & &
     \left.
     \begin{array}{ll}
     \gzbar(\mmz) &\!\!= 0.55557
      -0.00042\,\textfrac{\alpha_s' -0.1218}{0.0038}
       \pm 0.00061
      \\[1mm]
     \sbar(\mmz)  &\!\!= 0.23065
      +0.00003\,\textfrac{\alpha_s' -0.1218}{0.0038}
       \pm 0.00024
     \end{array}
    \right\}\,\,
   \rho_{\rm corr} = 0.24,
   \\ 
   & & \quad
   \chi^2_{\rm min} =  15.4
         +\biggl(\frac{\alpha_s' -0.1218}{0.0038}\biggr)^2
         +\biggl(\frac{\delb+0.0051}{0.0028}\biggr)^2\,. \\ \nonumber
   \label{fitofgzbsbchisq}
  \end{eqnarray}
  \end{subequations}
with (d.o.f) = 11.
The above result can be expressed in terms of the $S$ and 
$T$-parameters as follows;
  \bsub
  \begin{eqnarray}
     & &
     \left.
     \begin{array}{ll}
     S' &\!\!= -0.29
      -0.059\,\textfrac{\alpha_s' -0.1218}{0.0038}
       \pm 0.13
      \\[1mm]
     T  &\!\!= \hph 0.69 
      -0.102\,\textfrac{\alpha_s' -0.1218}{0.0038}
       \pm 0.15
     \end{array}
    \right\}\,\,
   \rho_{\rm corr} = 0.87,
   \\ 
   & & \quad
   \chi^2_{\rm min} =  15.4
         +\biggl(\frac{\alpha_s' -0.1218}{0.0038}\biggr)^2
         +\biggl(\frac{\delb+0.0051}{0.0028}\biggr)^2\,. \\ \nonumber
  \end{eqnarray}
\label{eq:zpole}
  \end{subequations}
\\
Here the combination $S' = S -0.72 \delta_\alpha$ is determined 
from the $\sbar(m_Z^2)$ measurement~\cite{hhm}.
It is slightly different from the combination 
$S''$ of Eq.~(\ref{eq:s_double_prime})  which is constrained 
from the $\sbar(0)$ measurement at low-energies.
The difference coms from the uncertainty in the hadronic corrections 
to the running of the effective 
charge $\ebar(q^2)/\sbar(q^2)$ between $q^2 = m_Z^2$ and 
$q^2 = 0$; see Appendix B of Ref.~\cite{hhkm}. 
The result is shown in Fig.~\ref{stu_fit} for $\delta_\alpha = 
0.03$~\cite{eidelman} and $\alpha_{s}'= 0.1218$. 
We find that the $Z$-pole results (\ref{eq:zpole}) are consistent 
with the SM predictions for $\xt = \xh = 0$, 
$(S,T)=(-0.23,0.88)$, whereas the results of the 
\lenc experiments  (\ref{eq:sum_lenc}) give slightly smaller $S$ and $T$. 

We can combine all the \lenc data with 
the above LEP/SLC data, and find for $\delta_\alpha = 0.03,$ 
  \begin{subequations}
  \begin{eqnarray}
     & &
     \left.
     \begin{array}{ll}
     S &\!\!= -0.33
      -0.051\,\textfrac{\alpha_s' -0.1227}{0.0037}
       \pm 0.13
      \\[1mm]
     T  &\!\!= \hph 0.61
      -0.089\,\textfrac{\alpha_s' -0.1227}{0.0037}
       \pm 0.14
     \end{array}
    \right\}\,\,
   \rho_{\rm corr} = 0.86,
   \\ 
   & & \quad
   \chi^2_{\rm min} =  20.8
         +\biggl(\frac{\alpha_s' -0.1227}{0.0037}\biggr)^2
         +\biggl(\frac{\delb+0.0051}{0.0028}\biggr)^2\,, 
  \end{eqnarray}
  \end{subequations}
with (d.o.f) = (15(L.E.N.C.)+13(LEP/SLC)$)-2 = 26$.
The combined fit does not change significantly from the previous 
result of Ref.~\cite{hhm} mainly because of the dominant role of 
the $Z$-pole data.

\clean
\section{ Constraints on contact terms}
In this section, we study the constraints on the contact interactions 
from all the \lenc data. 
\subsection{ Constraints on general contact interactions  }
In the general framework of the contact interactions, each \lenc 
experiment constrains certain combinations of the contact terms. 

From the SLAC $e$D experiments, two combinations of the model 
independent parameters $C_{1q}^e$ and $C_{2q}^e$ are constrained.  
Assuming the SM prediction (\ref{eq:slac_sm2}), 
the experimental data (\ref{eq:ed_exp}) 
lead the following constraints;
\bea
\begin{array}{l}
	\left.
	\begin{array}{l}
	\Delta( 2C_{1u}^e-C_{1d}^e) = \hph 0.217 \pm 0.26 \\
	\vsk{0.2}
	\Delta( 2C_{2u}^e-C_{2d}^e) = -0.765 \pm 1.23 \\
	\end{array}
	\right \}~~~~ 
	\rho_{\rm corr} = -0.975.
\end{array}
\label{eq:slac_contact}
\eea

Here and in the following, we adopt the SM predictions for 
$\mt = 175\ {\rm GeV}$ and $\mh = 100\ {\rm GeV}$ (and $\delta_\alpha = 
0.03$~\cite{eidelman}). 
Effects of small changes in the SM predictions for different 
values of $\mt$ and $\mh$ (or for $S$ and $T$) can easily 
be accounted for by using the parametrizations given in section 3.

The CERN $\mu^\pm$C experiment is used to extract the model 
independent parameters $C_{2q}^\mu$ and $C_{3q}^\mu$.  
From the experimental data given in section 3, we find the 
following constraints on the linear combination of 
$\Delta C_{2q}^\mu$ and $\Delta C_{3q}^\mu$; 
\bea
\begin{array}{l}
	\left.
	\begin{array}{l}
	\Delta( 2C_{3u}^\mu-C_{3d}^\mu) = -1.51 \pm 4.9 \\
	\vsk{0.2}
	\Delta( 2C_{2u}^\mu-C_{2d}^\mu) = \hph 1.74 \pm 6.3 \\
	\end{array}
	\right \}~~~~ 
	\rho_{\rm corr} = -0.997. 
\end{array}
\eea

The Bates experiment on $e$C scattering constrains the following 
combination; 
\beq
\Delta (C_{1u}^e + C_{1d}^e) = 
0.0152 \pm 0.033.
\eeq

The Mainz experiment on $e$Be scattering constrains the combination 
\bea
\Delta A_{Mainz} &=& 
-2.73\Delta C_{1u}^e + 0.65\Delta C_{1d}^e 
-2.19\Delta C_{2u}^e + 2.03\Delta C_{2d}^e \\ \nonumber
&=&
-0.065 \pm 0.19. 
\eea

The two APV experiments give; 
\bea
\Delta Q_W(Cs) &=& 
	376 \Delta C_{1u}^e + 422 \Delta C_{1d}^e = 0.96 \pm 0.92,\\
\label{eq:cs} 
\Delta Q_W(Tl) &=& 
	572 \Delta C_{1u}^e + 658 \Delta C_{1d}^e 
		= 1.58 \pm 4.2.  
\label{eq:tl}
\eea

By combining the two sets of the $\nu_\mu$-$q$ scattering data, 
we find
\bea
\begin{array}{l}
	\left.
	\begin{array}{l}
	\Delta u_L = -0.0032 \pm 0.0169 \\
	\Delta u_R = -0.0084 \pm 0.0251 \\
	\Delta d_L = \hph 0.0020 \pm 0.0136 \\
	\Delta d_R = -0.0109 \pm 0.0631 \\
	\end{array}
	\right \} ~~~
	\rho_{\rm corr} = \left ( 
	\begin{array}{rrrr}
	1 &  0.385 & 0.954 & 0.412 \\
	  &      1  & 0.355& 0.841 \\
	  &         &     1  & 0.511 \\
	  &         &        &      1 
	\end{array} \right ).
\label{eq:nq}
\end{array}
\eea
The contact term contributions to the model-independent parameters 
$\Delta C_{iq}$ and $\Delta q_\alpha$ 
are found in Eqs.~(\ref{eq:c_def}) and (\ref{eq:q_def}), respectively.

From the $\nu_{\mu}$-$e$ scattering experiments, (\ref{eq:nu_e}), 
we find 
\bea
\begin{array}{l}
	\left.
	\begin{array}{l}
	\Delta M_{LL}^{\nu_\mu e} =
	\eta_{LL}^{\nu_\mu e}
	= -0.15 \pm 0.35 \\
	\vsk{0.2}
	\Delta M_{LR}^{\nu_\mu e} =
	\eta_{LR}^{\nu_\mu e}
	= -0.03 \pm 0.36
	\end{array}
	\right \} ~~
	\rho_{\rm corr} = 0.13, 
\end{array}
\eea
in units of TeV$^{-2}$.
The two pure-leptonic contact interactions are constrained directly 
by these experiments.

\subsection{ Constraints on \sm invariant contact interactions  }
Without any further assumptions, 
there are 108 lepton-quark contact couplings $\eta_{\alpha \beta}^{\ell q}$. 
In this subsection, we combine the individual constraints 
(\ref{eq:slac_contact}) $\sim$ (\ref{eq:nq}) by taking the following 
assumptions; \\
(i) {\it  lepton universality}
\begin{eqnarray}
\eta^{eq}_{\alpha \beta} = 
\eta^{\mu q}_{\alpha \beta} = 
\eta^{\tau q}_{\alpha \beta} \equiv 
\eta^{\ell q}_{\alpha \beta},  
\end{eqnarray}
(ii) {\it  quark universality}
\bsub
\begin{eqnarray}
\eta^{\ell u}_{\alpha \beta} &=&
\eta^{\ell c}_{\alpha \beta} = 
\eta^{\ell t}_{\alpha \beta}, \\
\eta^{\ell d}_{\alpha \beta} &=&
\eta^{\ell s}_{\alpha \beta} = 
\eta^{\ell b}_{\alpha \beta}, 
\end{eqnarray}
\esub
(iii) {\it  \sm invariance (for {\rm SU(2)$_L$} singlet exchange) }
\bsub
\begin{eqnarray}
\eta^{\ell u}_{\alpha L} &=&
\eta^{\ell d}_{\alpha L}, \\
\eta^{\nu_\ell q}_{L \beta} &=&
\eta^{\ell q}_{L \beta}. 
\end{eqnarray}
\esub
Six lepton-quark contact terms remain as independent parameters;
\begin{eqnarray}
\biggl\{ 
\eta^{\ell u}_{LL}, 
\eta^{\ell u}_{LR}, 
\eta^{\ell d}_{LR}, 
\eta^{\ell d}_{RR}, 
\eta^{\ell u}_{RL}, 
\eta^{\ell u}_{RR}
\biggr\}. 
\end{eqnarray}

Let us first examine how the above six contact interaction terms are 
constrained from each low-energy experiment. 
We express the model-independent parameters, $\Delta C_{iq}, 
\Delta q_\alpha$ and $\Delta M_{\alpha \beta}^{\ell q}$ 
in terms of $\eta_{\alpha \beta}^{\ell q}$ with 
the above assumptions. 
We find the following constraints on the various combinations of 
the contact terms: Here $\eta_{\alpha \beta}^{\ell q}$ is measured 
in units of ${\rm TeV}^{-2}$.

SLAC $e$D experiment:
\bea
\begin{array}{l}
-0.5 \eta_{LL}^{\ell u} -0.657 \eta_{LR}^{\ell u} 
+0.329 \eta_{LR}^{\ell d}+0.329 \eta_{RL}^{\ell u}
+ \eta_{RR}^{\ell u}-0.5 \eta_{RR}^{\ell d} \\
\vsk{0.2}
= -0.86 \pm 0.79.
\end{array}
\label{eq:slac_cont_linear}
\eea

CERN $\mu^\pm$C experiment:
\bea
\begin{array}{l}
0.062 \eta_{LL}^{\ell u} - 0.124 \eta_{LR}^{\ell u} 
+0.062 \eta_{LR}^{\ell d} -0.5 \eta_{RL}^{\ell u}
+ \eta_{RR}^{\ell u}-0.5 \eta_{RR}^{\ell d} \\
\vsk{0.2}
= 1.4 \pm 3.5.
\end{array}
\label{eq:cern_cont_linear}
\eea

Bates $e$C experiment:
\bea
\begin{array}{l}
\eta_{LL}^{\ell u} + 0.5 \eta_{LR}^{\ell u} 
+0.5 \eta_{LR}^{\ell d} - \eta_{RL}^{\ell u}
-0.5 \eta_{RR}^{\ell u}-0.5 \eta_{RR}^{\ell d} \\
\vsk{0.2}
= 0.25 \pm 0.54.
\end{array}
\eea

Mainz $e$Be experiment: 
\bea
\begin{array}{l}
-0.455 \eta_{LL}^{\ell u} -0.110 \eta_{LR}^{\ell u} 
-0.280 \eta_{LR}^{\ell d} + 0.390 \eta_{RL}^{\ell u}
+ \eta_{RR}^{\ell u}-0.545 \eta_{RR}^{\ell d} \\
\vsk{0.2}
= -0.44 \pm 1.27.
\end{array}
\eea

APV($Cs$):
\bea
\begin{array}{l}
\eta_{LL}^{\ell u} + 0.471 \eta_{LR}^{\ell u} 
+ 0.529 \eta_{LR}^{\ell d} - \eta_{RL}^{\ell u}
-0.471 \eta_{RR}^{\ell u}-0.529 \eta_{RR}^{\ell d} \\
\vsk{0.2}
= 0.040 \pm 0.038.
\end{array}
\label{eq:apv_1}
\eea

APV($Tl$):
\bea
\begin{array}{l}
\eta_{LL}^{\ell u} + 0.465 \eta_{LR}^{\ell u} 
+ 0.535 \eta_{LR}^{\ell d} - \eta_{RL}^{\ell u}
-0.465 \eta_{RR}^{\ell u}- 0.535 \eta_{RR}^{\ell d} \\
\vsk{0.2}
= 0.042 \pm 0.113.
\end{array}
\label{eq:apv_2}
\eea
In Eqs.~(\ref{eq:slac_cont_linear}) and (\ref{eq:cern_cont_linear}), 
looser constraints are dropped. 

From the result of $\nu_\mu$-$q$ experiments (Eq.(\ref{eq:nq})), 
the following three contact terms are constrained;
\bea
\begin{array}{l}
	\left.
	\begin{array}{l}
	\eta^{\ell u}_{LL} = -0.22 \pm 0.40 \\
	\eta^{\ell u}_{LR} = \hph 0.06 \pm 0.83 \\
	\eta^{\ell d}_{LR} = \hph 0.47 \pm 2.14 \\
	\end{array}
	\right \}, ~~~
	\rho_{\rm corr} = \left ( 
	\begin{array}{rrr}
	1 &  0.27 & 0.54 \\
	  &    1  & 0.89 \\
	  &       &   1  
	\end{array} \right ).
\end{array}
\label{eq:nq_cont_res}
\eea
where, the contact terms are given in units of TeV$^{-2}$.

By adding all data of low-energy lepton-quark experiments 
given in Eqs.~(\ref{eq:slac_contact})
$\sim$ (\ref{eq:nq}), we find
\footnote{
In the actual fit, we used the original data, 
Eqs.~(\ref{eq:fh_data}) and (\ref{eq:ccfr_data}), 
instead of Eq.~(\ref{eq:nq}) to retain accuracy. 
This explains that the d.o.f. in Eq.~(\ref{eq:chisq_contact}) 
is not 6 but 7. 
Eq.~(\ref{eq:nq}) should be read as approximate constraints
since the $\nu_\mu$-$q$ `data' expressed in terms of 
$u_\alpha$ and $d_\alpha$ slightly deviate from 
the Gaussian.}
\bsub
\bea 
	\left .
	\begin{array}{rcl@{\,}c@{\,}l}
\eta_{LL}^{\ell u} &=& -0.281    &\pm & 0.375 \\
\eta_{LR}^{\ell u} &=& -0.081    &\pm & 0.739 \\
\eta_{LR}^{\ell d} &=&\hph 0.02  &\pm & 1.43  \\
\eta_{RL}^{\ell u} &=& -2.47     &\pm & 4.00  \\
\eta_{RR}^{\ell u} &=& \hph 1.39 &\pm & 3.53  \\
\eta_{RR}^{\ell d} &=& \hph 2.76 &\pm & 4.43
	\end{array}
	\right \},  
\label{eq:result_contact}
\eea 
\bea
\rho_{\rm corr} &=& \left( 
	\begin{array}{rrrrrr}
1 & 0.14 & 0.45 & 0.12 & 0.05 & 0.09 \\
&1& 0.86 & 0.13 & 0.11 & 0.15 \\
&&1& 0.17& 0.10& 0.17 \\
&&&1& -0.95 & -0.94 \\
&&&&1& 0.97 \\
&&&&&    1
	\end{array}
	\right ),\\
\chi^2_{\rm min}/({\rm d.o.f}) &=& 2.8/(7), 
\label{eq:chisq_contact}
\eea 
\esub 
\\
in units of TeV$^{-2}$. 

We give below the three eigenvectors of the $(6\times 6)$ covariance 
matrix with the smallest errors:
\bsub
\bea 
\begin{array}{l}
 0.353 \eta_{LL}^{\ell u} + 0.167 \eta_{LR}^{\ell u} 
+0.185 \eta_{LR}^{\ell d} - 0.351 \eta_{RL}^{\ell u}
-0.165 \eta_{RR}^{\ell u} - 0.185 \eta_{RR}^{\ell d} \\
\vsk{0.2}
= 0.014 \pm 0.021
\end{array}
\label{eq:cont_res1}\\
\vsk{0.3}
\begin{array}{l}
 0.283 \eta_{LL}^{\ell u} + 0.394 \eta_{LR}^{\ell u} 
-0.382 \eta_{LR}^{\ell d} + 0.182 \eta_{RL}^{\ell u}
+0.044 \eta_{RR}^{\ell u} + 0.129 \eta_{RR}^{\ell d} \\
\vsk{0.2}
= -0.15 \pm 0.25 
\end{array}
\label{eq:cont_res2}\\
\vsk{0.3}
\begin{array}{l}
  0.775 \eta_{LL}^{\ell u}- 0.726 \eta_{LR}^{\ell u} 
+ 0.009 \eta_{LR}^{\ell d} + 0.288 \eta_{RL}^{\ell u}
+ 0.270 \eta_{RR}^{\ell u} + 0.047 \eta_{RR}^{\ell d} \\
\vsk{0.2}
= -0.36 \pm 0.45
\end{array}
\eea 
\label{eq:cont_res3}
\esub
\\
The most accurate constraint (\ref{eq:cont_res1}) is essentially 
follows from the APV measurements (\ref{eq:apv_1}) and (\ref{eq:apv_2}). 
The second most accurate constraint (\ref{eq:cont_res2}) is essentially 
obtained by the $\nu_\mu$-$q$ scattering data (\ref{eq:nq}) 
or (\ref{eq:nq_cont_res}).
Finally the $\nu_\mu$-$e$ scattering experiments constrain the 
pure-leptonic contact interactions $\eta_{\alpha \beta}^{\ell \ell}$. 
We find in units of TeV$^{-2}$; 
\bea
\begin{array}{l}
	\left.
	\begin{array}{l}
	\eta_{LL}^{\ell \ell}
	= -0.15 \pm 0.35 \\
	\vsk{0.2}
	\eta_{LR}^{\ell \ell}
	= -0.03 \pm 0.36
	\end{array}
	\right \} ~~
	\rho_{\rm corr} = 0.13.
\end{array}
\label{eq:ne_contact}
\eea
As a reference, the minimal SM (all $\eta_{\alpha \beta}^{ff'} = 0$) 
gives an excellent fit to all the data 
\beq
\chi^2_{\rm SM}/(\rm d.o.f) = 6.9/(15).
\eeq
Therefore, low-energy electroweak experiments do not require 
any new interactions.
\clean	
\section{Discussion}
In this paper, we have studied the constraints on the 
four-Fermi contact interactions from low-energy electroweak 
experiments. 
From the polarization/charge asymmetry measurement of charged 
lepton-nucleus scattering experiments, atomic parity violation 
experiments, and from neutrino-quark scattering experiments, 
constraints on the lepton-quark 
contact interactions were obtained, while neutrino-electron 
scattering experiments constrain the lepton-lepton contact 
interactions. 
By assuming the flavor universality and the \sm gauge invariance 
of the contact interactions, those constraints are parametrized 
conveniently as the constraint (\ref{eq:chisq_contact}) 
for the six lepton-quark contact terms and (\ref{eq:ne_contact})
for the two pure-leptonic contact terms. 
\begin{figure}[b]
\begin{center}
\leavevmode\psfig{figure=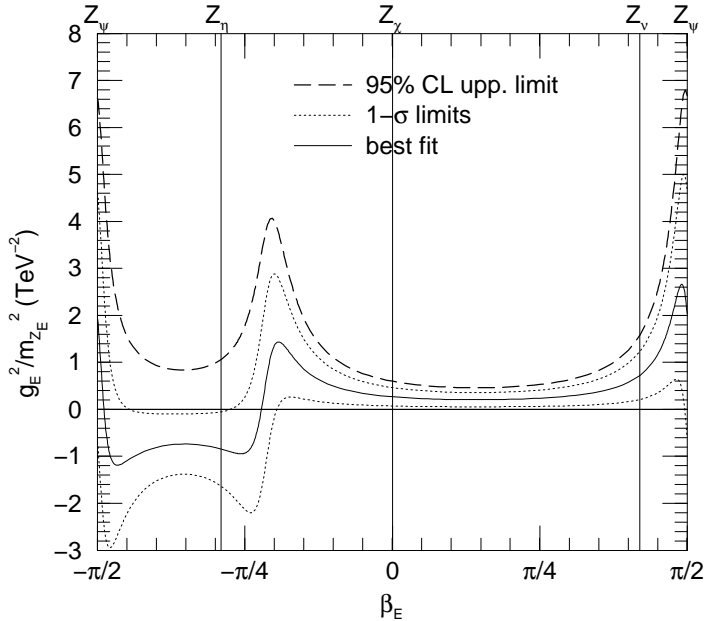,height=8.3cm}
\end{center}
\vspace{-4mm}
\caption{
Constraints on $g_E^2/m_{Z_E}^2$ from the low-energy electroweak 
measurements. 
The 1-$\sigma$ allowed region and the 95\% CL upper bounds are given 
for the $Z_E$ models that are characterized by the mixing angle 
$\beta_E$; see Eq.~(\protect\ref{eq:extra_z})}
\label{zcoupling}
\end{figure}
\begin{table}[b]
\caption{The hypercharge $Y$ and the extra U(1)$_E$ charge $Q_E$ of 
the left-handed quarks and leptons in $Z_\chi, Z_\psi,Z_\eta$ and 
$Z_\nu$ models. }
\begin{center}
\begin{tabular}{cccccc}  \hline \hline \\ [-0.4cm]
\vsp{0.2}
field & $Y$ & $\sqrt{24}Q_\chi$ & $\sqrt{72/5}Q_\psi$ & 
$Q_\eta$ & $Q_\nu$ \\ \hline \\
\vsp{0.2}
$\nu, e$ & $-\frac{1}{2}$& +3 & $+1$& $+\frac{1}{6}$& 
$+\sqrt{\frac{1}{6}}$\\
\vsp{0.2}
$\nu^c$  & $0$ & $-5$ & $+1$& $-\frac{5}{6}$& 0\\
\vsp{0.2}
$e^c$    & $+1 $ & $-1$& $+1$& $-\frac{1}{3}$& $+\sqrt{\frac{1}{24}}$\\
\hline \\
\vsp{0.2}
$u, d$   & $+\frac{1}{6}$ & $-1$& $+1$ & $-\frac{1}{3}$
& $+\sqrt{\frac{1}{24}}$\\
\vsp{0.2}
$u^c$    & $-\frac{2}{3}$ & $-1$& $+1$& $-\frac{1}{3}$ 
&$+\sqrt{\frac{1}{24}}$ \\
\vsp{0.2}
$d^c$    & $+\frac{1}{3}$ & $+3$& $+1$& $+\frac{1}{6}$& 
$+\sqrt{\frac{1}{6}}$ 
\\ \hline \hline 
\end{tabular} \\
\vsp{0.3}
\end{center}
\end{table}
%
\begin{table}[t]
\caption{
Constraints on $g_E^2/m_{Z_E}^2$ and $m_{Z_E}$ from the low-energy electroweak 
experiments for the four representative extra $Z$-boson models. 
We assumed $g_E^2 = g_Y^2 = 0.1270$ to obtain the 95\% CL lower limits of $m_{Z_E}$.
}
\begin{center}
\begin{tabular}{|c|c|c|c|c|c|} \hline \hline 
$Z_E^{\frac{}{}}$ &
$\beta_E$ & $g_E^2/m_{Z_E}^2$ & $\chi^2_{\rm min}$ &
\multicolumn{2}{c|}
{95\% CL limits} \\ \cline{5-6}
&& (TeV$^{-2}$)& & $g_E^2/m_{Z_E}^2$    & $m_{Z_E}$ 
\\  \hline 
$Z_\chi$ & 0      
& $0.25 \pm 0.20$ & 5.4 & 0.582 TeV$^{-2}$ & 470 GeV\\
$Z_\psi$ &$\pi/2$ & $2.0 \pm 2.5$ & 6.4 & 6.44 TeV$^{-2}$ & 140 GeV\\
$Z_\eta$ & $\tan^{-1} (-\sqrt{5/3}) $
& $-0.80 \pm 0.79$ & 6.9 & 1.08 TeV$^{-2}$ & 340 GeV\\
$Z_\nu$ & $ \tan^{-1} (\sqrt{15})$ 
&$0.69 \pm 0.51 $  & 5.2 & 1.54 TeV$^{-2}$ & 290 GeV\\ 
\hline \hline 
\end{tabular}\\
\vsp{0.3}
\end{center}
\end{table}

By using our result, it is easy to examine 
consequences of models of new physics that affect low-energy 
electroweak observables. 
As an example, we briefly study constraints on an 
extra $Z$-boson in ${\rm E}_6$ models. 
The models contain two additional new neutral gauge bosons, 
one is the SO(10) singlet $Z_\psi$, and the other one is the SU(5) 
singlet $Z_\chi$. 
In general the two gauge bosons are mixed, and the lighter one $Z_E$ 
is given by the following linear combination;
\bea
Z_E &=& Z_\chi \cos \beta_E + Z_\psi \sin \beta_E.
\label{eq:extra_z}
\eea
Mixing angle 
$\beta_E = 0, \pi/2, \tan^{-1}(-\sqrt{5/3})$ and $\tan^{-1}(\sqrt{15})$ 
correspond to $Z_\chi, Z_\psi, Z_\eta$ and $Z_\nu$ models, respectively. 
Following the Particle Data Group notation~\cite{PDG}, the 
hypercharge $Y$ and the extra U(1)$_E$ charge $Q_E$ of the 
left-handed quarks and leptons are given by;
Then, expressing the contact terms by U(1)$_E$ gauge coupling $g_E$ 
and the extra gauge boson mass $m_{Z_E}$ as,  
\bsub
\bea
\eta_{\alpha \beta}^{ff'} &=& -\frac{g_\alpha^f g_\beta^{f'}}{m_{Z_E}^2}, \\
g_L^f &=& \hph g_E Q_E^f, \\
g_R^f &=& -g_E Q_E^{f^C},
\eea
\esub
we can find the constraints on each extra $Z$-boson model. 
We show in Fig.~\ref{zcoupling} the constraints on 
$g_E^2/m_{Z_E}^2$ for each model parametrized by 
the mixing angle $\beta_E$.  
\begin{figure}[b]
\begin{center}
\leavevmode\psfig{figure=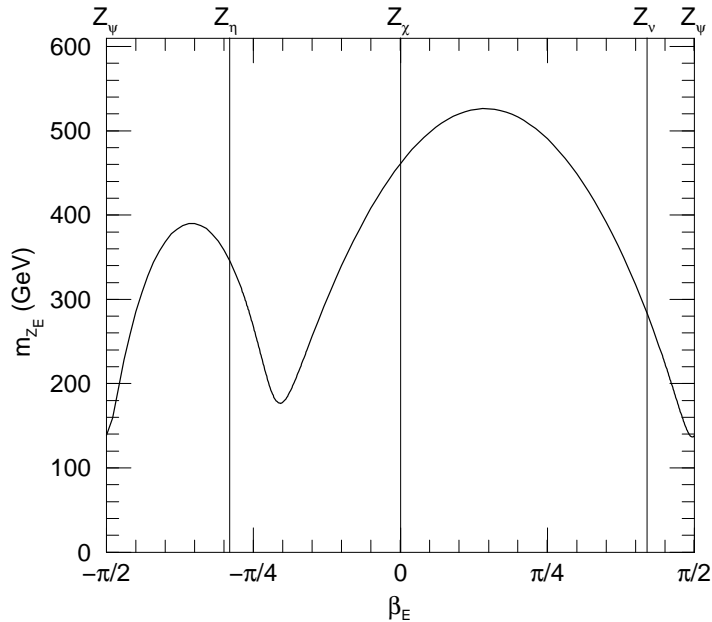,height=8.3cm}
\end{center}
\vspace{-4mm}
\caption{
95\% CL lower limits of the extra $Z$ boson mass $m_{Z_E}$ as 
a function of the mixing angle $\beta_E$.}
\vspace{3mm}
\label{zmass}
\end{figure}
%
In the figure, the region of $g_E^2/m_{Z_E}^2 < 0$ is unphysical. 
The 95\% CL upper bounds are obtained under the constraint 
$g_E^2/m_{Z_E}^2 > 0$. 
The reason for the appearance of the peak at $|\beta_E| = \pi/2$ 
($Z_\psi$ model) can be easily understood as follows; 
in the $Z_\psi$ model, all left-handed fermions have the same 
U(1)$_E$ charge, and hence the couplings are Parity conserving, 
which makes the most rigorous constraints from APV experiments useless.
The set of six couplings, $\eta_{\alpha \beta}^{\ell q}$ , 
is also less constrained by APV measurements at $\beta_E \simeq \pi/5$,  
thus the another peak is appeared.
To estimate the limit on the extra $Z$-boson mass, 
the extra U(1)$_E$ coupling $g_E$ should be fixed. 
Assuming $g_E^2 = g_Y^2 = \ebar(m_Z^2)/\cbar(m_Z^2) = 0.1270$, 
we show the 95\% CL lower limit on 
the extra $Z$-boson mass in Fig.~\ref{zmass}.

Constraints on $g_E^2/m_{Z_E}^2$ from the low-energy electroweak 
experiments for the four representative extra $Z$-boson models 
are listed in Table 2. 
The 95\% CL upper limits of $g_E^2/m_{Z_E}^2$ and 
the 95\% CL lower limits of $m_{Z_E}$ for $g_E^2 = g_Y^2 = 0.1270$ 
are also given in the Table 2. 
As compared to Ref.~\cite{lang}, $Z_\chi$ and $Z_\eta$ models are 
more severely constrained by updated low-energy electroweak 
experiments. 
%

\section*{Acknowledgement}

The authors wish to thank V. Barger, K. Cheung, D. Haidt, K. Tokushuku 
and D. Zeppenfeld for stimulating discussions. 
The work of K.H. is supported in part by the JSPS-NSF Joint Research 
Project. 
The work of G.C.C. and S.M. is supported in part by Grant-in-Aid 
for Scientific Research from the Ministry of Education, Science
and Culture of Japan. 

\newpage

\end{document}